\documentclass[%
 aip,
rsi,%
 amsmath,amssymb,
 reprint,%
]{revtex4-1}

\usepackage{graphicx}
\usepackage{dcolumn}
\usepackage{bm}

\begin{document}

\preprint{AIP/123-QED}

\title{Nonlinear Frequency Shift in Raman Backscattering and its Implications for Plasma Diagnostics}

\affiliation{ 
Plasma Physics Division, Naval Research Laboratory, Washington, DC 20375, USA
}
\author{D. Kaganovich}
\email{dmitri.kaganovich@nrl.navy.mil}
\author{B. Hafizi}
\author{J. P. Palastro}
\author{A. Ting}
\altaffiliation{Also at Research Support Instruments, Lanham, Maryland 20706, USA}
\author{M. H. Helle}
\author{Y.-H. Chen}
\altaffiliation{Also at Research Support Instruments, Lanham, Maryland 20706, USA}
\author{T. G. Jones}
\author{D. F. Gordon}

\date{\today}

\begin{abstract}
Raman backscattered radiation of intense laser pulses in plasma is investigated for a wide range of intensities relevant to laser wakefield acceleration. The weakly nonlinear dispersion relation for Raman backscattering predicts an intensity and density dependent frequency shift that is opposite to that suggested by a simple relativistic consideration. This observation has been benchmarked against experimental results, providing a novel diagnostic for laser-plasma interactions. 
\end{abstract}

\maketitle

\section{Introduction}

The stimulated Raman scattering instability in plasma occurs when an incident pump laser pulse scatters from an electron plasma wave\cite{kruer1988, mc1992, antonsen1993, gordon2002, esarey2009, palastro2011}. The resulting scattered electromagnetic waves are downshifted (Stokes) and upshifted (anti-Stokes) by the plasma frequency. For direct backscattering, the growth rate is maximized, and the ponderomotive force resulting from the beating of the Stokes and pump waves reinforce the plasma wave. This, in turn, increases the energy scattered into the Stokes sideband. The feedback process continues until the Stokes wave itself becomes unstable or another nonlinear process takes over \cite{Yin2008,palastro2009}.

Amplification of the Raman backscattered wave is physically equivalent to one of the operating regimes of the free electron laser.  In either case relativistic analysis of the scattering process shows that the frequency is a complex-valued number with both real and  imaginary parts. The imaginary part is of course connected with growth of the backscattered wave.  The real part results in a pump amplitude dependent frequency shift that is physically distinct from the shift associated with the effective electron mass increase due to relativity.

The wavelength of the back scattered radiation is commonly used to measure the plasma density in gas jets\cite{perry1992, krushelnick1995} and plasma channels\cite{jones2002} used in laser wakefield acceleration (LWFA) experiments. Only a limited number of experiments, however, have attempted to compare the Raman based plasma density to independent methods like interferometry\cite{ping2003}. At the same time, the theory of Raman backscattering (RBS) is limited to intermediate laser intensities, below those acheivable by modern ultra-intense lasers. This deficiency calls into question plasma density retrieval from the measured RBS spectrum at intensities characteristic of laser wakefield accelerators.

At high laser intensities the RBS frequency shift is a function of both plasma density and laser intensity.  This places a limitation on the use of Raman backscatter spectroscopy as a plasma density diagnostic. On the other hand, detailed understanding of the scattering process can provide useful information about laser-plasma interactions. For example, it is shown below that Raman backscattered radiation can be used to estimate the laser intensity inside a known density plasma target, provide information on the self-focusing conditions, or determine the optimal plasma density for quasi-linear wakefield acceleration. In this paper we present a detailed experimental study of nonlinear Raman backscattering occuring in conditions relevant to laser wakefield acceleration. Saturation of the RBS frequency shift occurs concomitantly with the acceleration of MeV class electrons. The observations are shown to be in close agreement with theoretical analysis in the weakly nonlinear regime. 

\begin{figure*}[htb]
\begin{center}
\includegraphics[width=6.0in]{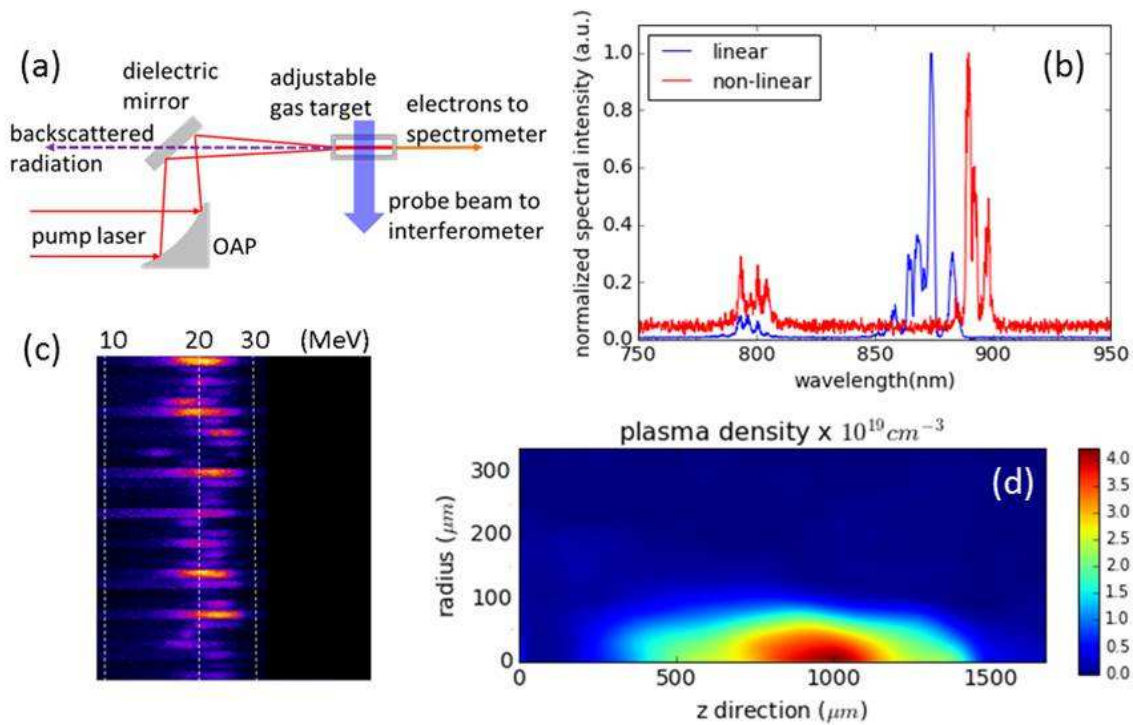}
\caption{(a) Experimental setup, OAP - off axis parabola, (b) typical RBS spectrum for linear and nonlinear case (see text for details), (c) energetic electron spectrum from 32 consecutive shots, and (d) reconstructed plasma density profile of the target similar to one used to obtain results in (b) and (c). Pump laser in (d) propagates from the right along the axis of symmetry.}
\label{fig1}
\end{center}
\end{figure*}

\section{Experimental}

Experimental studies of Raman backscattering were conducted using the LWFA facility at the Naval Research Laboratory \cite{gordon2013}. The goal was to measure the RBS frequency shift as a function of plasma density and pump laser intensity. The typical experimental arrangement is shown in Figure  \ref{fig1}a and discussed below. A linearly polarized, 50 fs compressed, 400 mJ, 8 TW peak power (on target), 800 nm central wavelength pump laser was focused by an F-number 10 off-axis parabolic mirror onto an adjustable gas target \cite{kaganovich2014} inside a vacuum chamber. A fully-ionized plasma column formed along the laser propagation direction. The laser focal radius was measured in vacuum with 86.5\% of the laser energy contained within a 5.1 micron spot. The gas target was attached to a 3-axis translation stage allowing adjustment of the focal point alignment. A small, frequency-doubled portion of the pump pulse was separated and used to probe the interaction region. This 400 nm probe beam was used to obtain the plasma density using folded wave, transverse interferometery a few picoseconds after the interaction. For plasma densities above $5\times10^{19} cm^{-3}$ the density was determined from neutral gas interferometry with the pump laser blocked before entering the interaction region. Helium gas was used in all experiments and it was fully ionized by the pump laser for the range of experimental intensities.

\begin{figure}[htb]
\begin{center}
\includegraphics[width=3.0in]{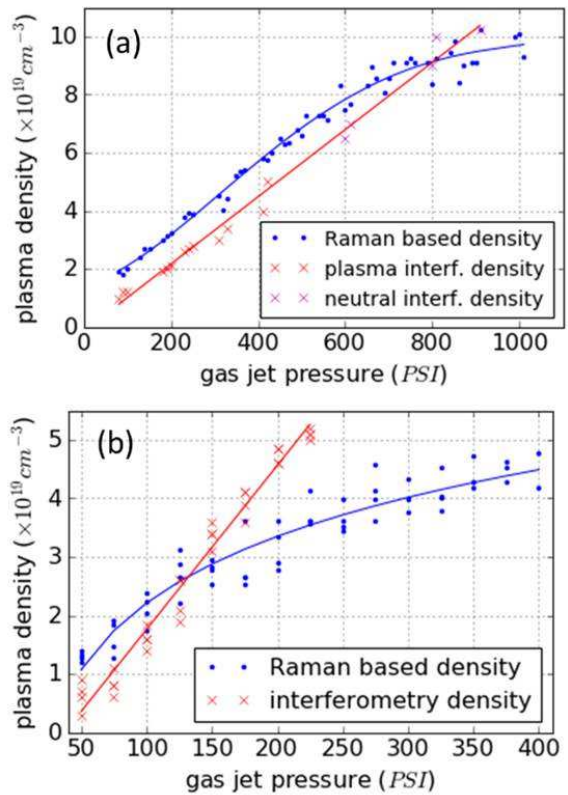}
\caption{Peak plasma density as a function of backing pressure in the gas jet. (a) Stable down-ramp acceleration conditions presented in Figure 1; (b) results from a 2 mm-long flat-top density gas jet with the laser focused in the middle. The plasma density was reconstructed from interferometry measurements and from the linear approximation $n_R$ for the RBS frequency shift in Eq. \ref{eq2}. Each point on the graph corresponds to one independent measurement, solid lines are plotted to provide a visual guide of data trends. Large fluctuations in the interferometry data at low pressures in (b) are due to small fringe shifts.}
\label{fig2}
\end{center}
\end{figure}

The Raman backscatter radiation was collected by an F-number 20 lens through a tunable laser line mirror (CVI TLM1-800-45). The mirror served as a $\approx \pm$25 nm bandwidth notch filter to reduce the background level of the scattered 800 nm pump. The collected back-reflected light was focused on the slit of an imaging spectrometer, coupled to a 16 bit linear array CCD sensor and to a picosecond-resolution streak camera.  A typical output is shown in Figure \ref{fig1}b. The streak camera was used to distinguish between the RBS radiation that arrived first, and stronger, but later arriving, light reflected off the walls of the vacuum chamber and from a light block positioned between the gas target and the electron spectrometer. These reflections arrived within a nanosecond-scale delay and were eliminated by proper arrangement of the light block and target positioning.

The plasma density profile was adjusted by varying the gas jet backing pressure, nozzle tilt and nozzle geometry \cite{kaganovich2014}. Experimental results were collected using a variety of gas targets, but for most of the experiments presented here a 0.75 mm-internal-diameter, 6 mm-long metal pipe was used as the nozzle. The nozzle was tilted 15 degrees off the normal to the pump laser propagation direction (from right to left in Figure \ref{fig1}d) and provided the density profile shown in Figure 1d as reconstructed from interferometry. The pump focal position was moved towards the elongated (due to the tilt) down-ramp side of the density profile (z-position between 600 and 700 $\mu m$ in Figure \ref{fig1}d) to create a gradient-assisted electron injector \cite{gordon2013, brantov2008} capable of delivering stable 20 MeV electron bunches (Figure \ref{fig1}c). The electron energy spectrum was acquired by an electron spectrometer described elsewhere \cite{helle2010}. The pump laser intensity was varied by detuning the laser compressor, calibrated using high-dynamic-range third-order polarization frequency-resolved optical gating \cite{kaganovich2013}. The laser pulse was stretched over the range 50 fs - 3 ps in both positive and negative chirp directions. The results presented below were found to be independent of the chirp direction.

In the first set of experiments the peak plasma density measured with interferometry was compared to the density calculated from the frequency shift of the RBS radiation at peak pump power. Assuming a linear frequency-matching condition for the backscattered Stokes wave, 
\begin{align}
\omega_1 = \omega_0 - \omega_p
\label{eq1}
\end{align}
an effective plasma density $n_{R}$ can be calculated from the plasma frequency $\omega_p=(4\pi q^{2}n_{R}/m)^{1/2}$, where q and m denote electron charge and mass. The RBS frequency $\omega_1$ is determined by $2\pi c/\lambda_1$, where $\lambda_1$ is the measured backscattered wavelength farthest from $\lambda_{0}=800 nm$ (see Fig. \ref{fig1}b for a typical spectrum). For calculation purposes we rewrite Eq. \ref{eq1} as
\begin{align}
n_{R}(10^{19} cm^{-3})=1.11\times 10^{8}\left(\frac{1}{\lambda_{0}(nm)}-\frac{1}{\lambda_{1}(nm)}\right)
\label{eq2}
\end{align}

In LWFA, careful tailoring of plasma density is crucial to effective electron acceleration by the laser driven plasma waves.  Comparison of plasma density measurements using interferometry and Raman frequency shift was routinely performed in the experiments to improve the accuracy of plasma density measurements. However, during careful examination of results from several sets of different LWFA experimental conditions, consistent discrepancies were seen between the two different density measurement techniques. Figure \ref{fig2} shows  plasma density measurements based on interferometry (red crosses) and on Eq. \ref{eq2} (blue dots) as functions of the backing pressure for two different plasma density profiles. Figure \ref{fig2}a shows results from the down-ramp density accelerator described above and displayed in Figure \ref{fig1}, while Figure \ref{fig2}b shows results from a 2 mm long, slit-shaped, flat-top, 0.5 mm-ramp gas jet with the pump laser focused in the middle. Both jets generated LWFA electrons but in case of the down-ramp accelerator stable beams of monoenergetic electrons (Figure 1c) were generated at high plasma densities (between 7 and 8$\times10^{19} cm^{-3}$). In the flat-top density profile forward accelerated electrons started to appear at a lower plasma density of 2.5 - 3 $\times10^{19} cm^{-3}$, but acceleration was unstable with shot-to-shot fluctuations and a broad electron energy spectrum in most  shots.

While Figs. \ref{fig2}a and \ref{fig2}b appear different, they are similar in several important respects: i) the interferometry measurements represent the “true” plasma density and the gas density of both gas jets and nozzles is a linear function of the backing pressure; ii) the plasma density, $n_R$, calculated from Eq. \ref{eq2} at low densities is also a linear function of the gas pressure, but is off-set to a higher plasma density indicating a larger shift of the Raman frequency than expected from the interferometry measured density; iii) at higher plasma densities the saturation observed in the Raman based curves coincides with the appearance of forward-accelerated electrons.
 
The saturation of the Raman based curve can be ascribed to the disturbance or destruction of the Raman plasma wave by cavitation and bubble formation wave breaking, and phase-mixing occurring in nonlinear LWFA \cite{helle2010}. In these conditions, the Raman driven plasma wave no longer has a quiescient plasma background from which to grow.  The laser power in this  case significantly exceeds the critical power (see discussion section), causing catastrophic self-focusing over relatively short interaction distances within the gas jets. The bubble regime of LWFA relies on laser self-focusing in order to reach the necessary intensities \cite{gordon2013, helle2010}. At increasing gas pressures laser self-focusing starts earlier along the laser propagation path, and eventually takes place before the peak density region of the gas target. The onset of strong self-focusing, followed by bubble formation, terminates the Raman backscattering leading to the saturation observed in Figure \ref{fig2}. The same arguments can be used to explain why the two curves cross at different pressures in (a) and (b): the laser propagates a longer distance and with higher intensity (the geometric focus is located at the peak of the plasma density) in (b) and, as a result, experiences self-focusing at a lower density than in (a).

\begin{figure}[htb]
\begin{center}
\includegraphics[width=3.0in]{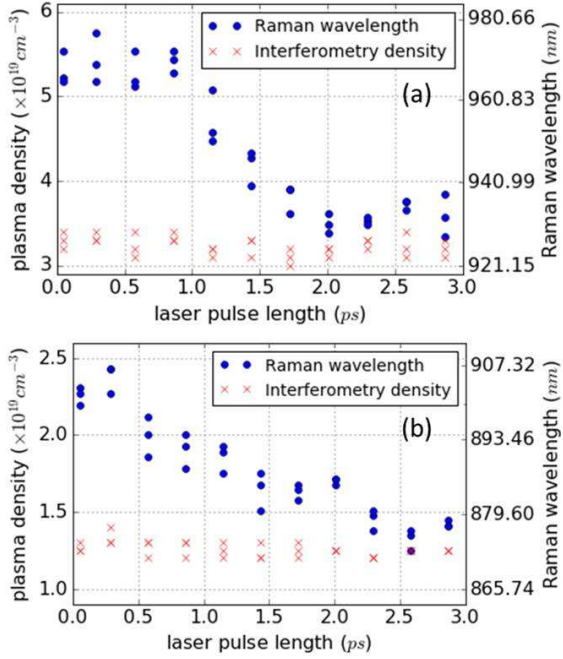}
\caption{Peak plasma density as function of the pump laser duration from the down-ramp acceleration conditions presented in Figure \ref{fig1} for high density (a) and low density (b) cases, both from below the linear section of the RBS curve in Figure \ref{fig2}a. Vertical axes are connected by Eq. \ref{eq2} and are valid for both sets of data points. Scale of the right vertical axis is logarithmic. Each point on the graph corresponds to one independent measurement.}
\label{fig3}
\end{center}
\end{figure}

The deviation of the Stokes frequency from the linear matching condition in Eq. \ref{eq2} was further investigated by varying the pump duration (and the intensity) at fixed plasma densities. The results are displayed in Figure \ref{fig3} for two different plasma densities. The densities were chosen to be below self-focusing conditions, determined from the linear section of the $n_R$ curve in Figure \ref{fig2}a. The  interferometry density is, of course, independent of the laser intensity. In contrast, the effective density $n_R$ derived from the frequency of the backscattered light (Eq. \ref{eq2}) changes substantially with the pulse duration and approaches the actual plasma density only when the laser pulse is stretched. Figure \ref{fig1}b shows typical spectral shift of the RBS radiation from Figure \ref{fig3}b run; the only difference between the linear and nonlinear shots is the laser pulse duration. 

A corresponding decrease in the laser intensity was also achieved by decreasing the laser energy, while keeping the laser pulse duration fixed at 50 fs. Similar to Figures \ref{fig1}b and \ref{fig3}, transitions of the RBS wavelength were observed, but the back-reflected signal dropped below the noise level when the pump laser energy was decreased below 20 mJ. This placed a lower limit on the scanned laser intensities.

\section{Analysis}

\subsection{Dispersion Relation}

The stimulated Raman instability in plasma is mediated by plasma oscillations $(\omega, {\bf k})$, $\omega \approx \omega_p=(4\pi q^2 n_0 / m)^{1/2}$ and results from the scattering of an electromagnetic pump wave $(\omega_0, {\bf k_0})$ into Stokes and anti-Stokes electromagnetic waves $(\omega_0 \pm \omega, {\bf k_0} \pm {\bf k})$ \cite{kruer1988, mc1992, antonsen1993, gordon2002, esarey2009, palastro2011}. For direct forward scattering  ${\bf k} \approx {\omega_p/c}$ and both the upshifted and downshifted light waves can be excited.  The phase velocity of the plasma wave is nearly equal to $c$, and this process is known to be important in generation of energetic electrons in laser wakefield acceleration \cite{esarey2009}. Of interest here is the direct backscattering process, ${\bf k} \approx 2{\bf k_0}$, where the growth rate is maximized.  

The wave equation for the electric field ${\bf E}({\bf r},t)$ associated with the pump pulse and the Raman backscatter, correct to order $|{\bf E}|^3$, is \cite{esarey1997}
\begin{align}
\left(\nabla^{2}-\frac{1}{c^2}\frac{\partial ^2}{\partial t^2}\right){\bf E}=\frac{\omega^{2}_{p}}{c^2}\left(1+\frac{\delta n}{n_0}-\frac{|a|^2}{2}\right){\bf E}
\label{eq3}
\end{align}
where $n_0$ is the unperturbed plasma density, $\delta n$ is the plasma density perturbation associated with the Raman plasma wave, $a=\left(q/mc\omega_0\right)\left<{\bf E}\cdot{\bf E}\right>^{1/2}$ is the magnitude of the electron oscillation momentum normalized to $mc$ and the brackets denote a time average. Here we distinguish between the Raman plasma wave and the wakefield plasma wave; the former having a wavenumber $k\sim2k_0$ and the latter $k \sim \omega_p /c$. The last two terms on the right hand side of Eq. \ref{eq3} represent, respectively, the Raman plasma wave and the relativistic mass correction.

The electric field associated with the Raman plasma wave ${\bf E}_{p}({\bf r},t)$, correct to lowest order in $a^2$, is  
\begin{align}
\left(\frac{\partial ^2}{\partial t^2}+\omega^{2}_{p}\right){\bf E}_{p}=\frac{mc^2}{q}\frac{\omega^{2}_{p}}{2}\nabla |a|^2
\label{eq4}
\end{align}
with the associated density perturbation  given by
\begin{align}
\frac{\delta n}{n_0}=\frac{q}{m\omega^{2}_{p}}\nabla \cdot {\bf E}_p
\label{eq5}
\end{align}
The normalized vector potential $a=a_{0}+a_{1}$, written as the sum of terms associated with the laser and the backscatter, is expressed as rapidly varying carrier waves modulating slowly varying envelopes: 
\begin{align}
a_{0}=\frac{1}{2}\hat{a}_0 e^{i(k_0 z-\omega_0 t)}+c.c.
\nonumber
\end{align}
and
\begin{align}
a_{1}=\frac{1}{2}\hat{a}_1 e^{i(k_1 z+\omega_1 t)}+c.c.
\nonumber
\end{align}
where $\omega_0 = ck_0$, $\omega_1 = ck_1$ and $\omega_0 - \omega_1 \approx \omega_p$.
Similarly, the density perturbation is expressed as
\begin{align}
\delta n=\frac{1}{2} \delta \hat{n} e^{i[(k_{0}+k_{1})z-(\omega_0 -\omega_{1})t]}+c.c.
\nonumber
\end{align}

\begin{figure}[htb]
\begin{center}
\includegraphics[width=3.0in]{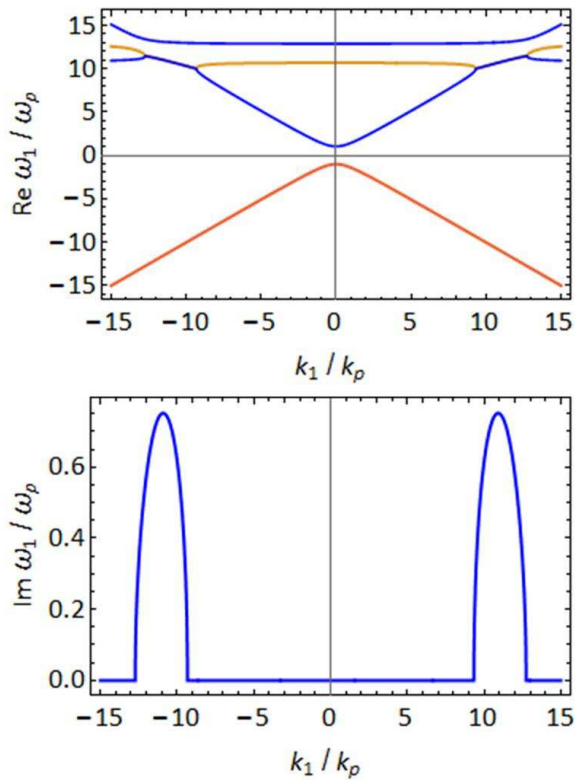}
\caption{Dispersion relation diagram describing Raman backscattering instability for plasma density $n_0 = 1.25\times 10^{19} cm^{-3}$ and laser intensity $I=4.2 \times 10^{17} W/cm^2$. The figure shows the real $Re(\omega_1)$ and imaginary $Im(\omega_1)$ parts of the frequency associated with Raman backscatter versus the wavenumber $k_1$ obtained from dispersion Eq. \ref{eq8}. The regions of instability are limited to the range of wavenumbers where an electromagnetic branch and a plasma branch coalesce into one temporally growing radiating mode.}
\label{fig4}
\end{center}
\end{figure}

Inserting the expression for the vector potentials and the perturbed density into Eqs. \ref{eq3} and \ref{eq4} one obtains the dispersion relation 
\begin{align}
\omega^{2}_{1} - \omega^{2}_{em} =\frac{\omega^{2}_{p}\hat{a}^{2}_{0}c^2 k^{2}_{0}}{(\omega_1 - \omega_0)^2 - \omega^{2}_{p}}
\label{eq6}
\end{align}
where
\begin{align}
\omega_{em} \equiv \sqrt{c^2 k^{2}_{1}+\omega^{2}_{p}\left(1-\frac{3}{4}\left|\hat{a}_{0}\right|^{2}\right)}
\label{eq7}
\end{align}
is the natural mode frequency with the wavenumber $k_1$ in the absence of coupling. Equation \ref{eq6} describes direct Raman backscatter (in the limit $a^2<1$).  A more general treatment, covering forward and backward Raman scattering, is given in Refs. [3] and [5]. To analyze Eq. \ref{eq6} it is instructive to re-write it in the form
\begin{align}
(\omega_1 - \omega_{em})(\omega_1 + \omega_{em})(\omega_1 - \omega_{+})(\omega_1 - \omega_{-})= \omega^{2}_{p}\hat{a}^{2}_{0}c^2 k^{2}_{0}
\label{eq8}
\end{align}
where $\omega_{\pm } \equiv \omega_0 \pm \omega_p$ are the frequencies associated with the anti-Stokes and Stokes modes. The real and imaginary parts of the solutions of Eq. \ref{eq8} are plotted in Figure \ref{fig4} for the experimental parameters and the longest laser pulse presented in Figure \ref{fig3}b.

Defining the frequency shift $\delta \omega = \omega_{-} - \omega_1$, and noting that $(\omega_1 + \omega_{em}) \approx 2\omega_{em}$, the dispersion relation simplifies to
\begin{align}
(\omega_0 - \omega_p - \delta \omega - \omega_{em})2\omega_{em}(\delta \omega +2\omega_p)\delta \omega \approx \omega^{2}_{p}\hat{a}^{2}_{0}c^2 k^{2}_{0}
\label{eq9}
\end{align}
In the weak pump limit $\hat{a}\ll 4\sqrt{\omega_p/\omega_0}$ Eq. \ref{eq9} can be solved to obtain
\begin{align}
\delta \omega = ick_0 \hat{a}_0 \sqrt{\frac{\omega_p}{4\omega_{em}}}\left[1-i\frac{ck_0 \hat{a}_0}{4\omega_p}\sqrt{\frac{\omega_p}{4\omega_{em}}}\right]
\label{eq10}
\end{align}
for the fastest growing, resonant mode $\omega_{em}=\omega_0 - \omega_p$. The latter can be solved along with the definition in Eq. \ref{eq7} to obtain the wavenumber $k_1$ of the resonant mode.

For practical applications, noting that the experiments and analysis apply to a linearly polarized laser beam, it is sufficient to employ an approximate form of Eq. \ref{eq10}
\begin{align}
\delta \omega \approx i\frac{{\hat{a}}_0}{2}\sqrt{\omega_0 \omega_p}\left[1-i\frac{{\hat{a}}_0}{8}\sqrt{\frac{\omega_0}{\omega_p}}\right]
\label{eq11}
\end{align}
The Raman side-bands move farther from the pump with increasing intensity, opposite what would be predicted by substituting the mass-corrected plasma frequency into the linear dispersion relation. By substituting $\hat{a}_0 = 8.6 \times 10^{-10} \lambda_{0}[\mu m] I^{1/2}_{0}[W/cm^2]$ and $\lambda_p=2\pi c/\omega_p$, we can express the wavelength of the Raman backscattered light in terms of experimentally-measurable parameters as
\begin{align}
\lambda_1 \approx \frac{\lambda_0}{1-\lambda_0 / \lambda_p - (\hat{a}_0/4)^2}
\label{eq12}
\end{align}
in the weak pump limit.

\subsection{Self-focusing and cavitation}

Detailed investigation of laser self-focusing is beyond the scope of this paper and is considered only in the context of the Raman backscattered spectrum observed experimentally in Figure \ref{fig2}. When the laser power $P_0=\pi r^{2}_{0} I_0/2$   is in excess of the critical power \cite{esarey2009}  $P_c \equiv 2c(mc^2 /e)^2(\omega_0/\omega_p)^2$  self-focusing due to relativistic effects takes place. It is assumed that the laser beam intensity $I_0$ is in the form of the fundamental Gaussian mode with spot radius $r_0$. Moreover, the ponderomotive force associated with the laser beam tends to expel electrons from regions of high intensity.  For sufficiently high power, cavitation and “bubble” formation result \cite{pukhov2002, esarey2009}. This is the regime of operation of most recent LWFA experiments as well as the experiments described herein (see Sect. II).

\section{Discussion}

The experiments described in Section II indicate that RBS is not generated from or beyond the axial location where electron acceleration is initiated.  The pump power was changed by varying the pulse duration. As the pulse length is shortened the laser power increases above $P_c$. Concomitant ponderomotive channeling expels electrons from regions of high intensity.  For the experimental parameters here ($r_0=5.1 \mu m$ and $n_0 = 3.3 \times 10^{19} cm^{-3}$) the critical power is $P_c \approx 9.2 \times 10^{11} W$ and, making use of the theory in Ref. [20], the electron density on axis vanishes (onset of blow-out regime) for $P_0/P_c \approx 1.42$. At the critical power, on-axis collapse and cavitation take place in a distance on the order of $400 \mu m$. It should be noted, however, that the analysis in Ref. [20] is for a uniform, homogeneous plasma and assumes that the laser beam maintains a Gaussian transverse profile throughout.

\begin{figure}[htb]
\begin{center}
\includegraphics[width=3.0in]{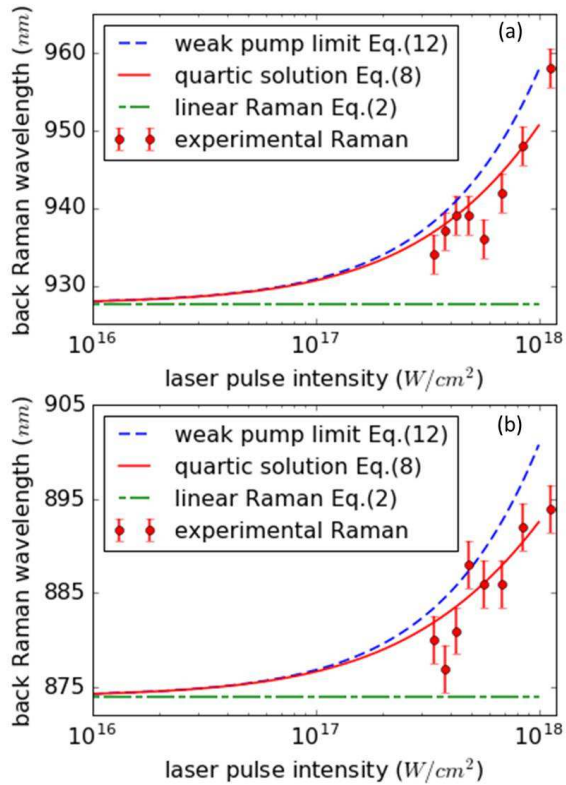}
\caption{Wavelength of fastest growing Raman backscatter light versus laser intensity.  High density example (a) with plasma density $n_0 = 3.3 \times 10^{19} cm^{-3}$, and low density (b) $n_0 = 1.25\times 10^{19} cm^{-3}$. Points refer to corresponding averaged experimental data in Figure \ref{fig3} and curves are from theory equations.}
\label{fig5}
\end{center}
\end{figure}

At lower laser intensities the RBS signal is clearly observed experimentally, but deviates strongly from the linear matching conditions of Eq. \ref{eq2} as intensity increases. Figure \ref{fig5} compares experimental results presented in Figure \ref{fig3} with the dispersion relation of Eq. \ref{eq8} and a practical formula valid in the weak pump limit of Eq. \ref{eq12}. The highest intensity experimental point displayed in Figures \ref{fig5}a and \ref{fig5}b still satisfies condition $\hat{a}_0 < 1$, so that the analytical solutions remain valid.  

Figure \ref{fig5} clearly demonstrates the inappropriateness of the linear approximation, Eqs. \ref{eq1} and \ref{eq2}, for evaluation of the plasma density at high laser intensities. On the other hand, no fitting parameters were used in our experimental-analytical comparison, making the agreement with the weakly nonlinear solution quite remarkable.

In the weak pump limit the real part of the frequency shift in Eq. \ref{eq10} closely matches the experimental data, and can be estimated by Eq. \ref{eq12}. This practical formula deviates somewhat from the experimental data and the analytical solution of Eq. \ref{eq8} at higher intensities, but is obviously a better alternative to the commonly used, intensity independent Eq. \ref{eq2}.

A recently proposed idea for a long guiding plasma channel \cite{kaganovich2015} might significantly improve performance of laser-plasma accelerators. At the same time it introduces a new challenge: diagnosing the guiding quality over a meter-scale length channel. In order to guarantee stable operation of the guide and to avoid unnecessary loss of pump energy, the laser spot size must match the plasma density profile \cite{ehrlich1998}. Additionally, optimizing the guiding performance in a resonant LWFA constrains the on-axis plasma density in the channel to a relatively narrow range \cite{hubbard2001}. Nonlinear and linear RBS can be used to measure both the matching radius of the channel and on-axis plasma density by time resolved (streak camera) analysis of the spectrum. A time resolved spectral plasma diagnostic in a channel was demonstrated in Ref. [11], however without employing  any nonlinear corrections. The dispersion relation, Eq. \ref{eq8}, can provide the on-axis plasma density in the channel for low pump intensities, where Raman shift is intensity independent. In addition it can indicate beam-spot size (and intensity) oscillations in a mismatched channel by showing temporally resolved spectral oscillations in the weakly nonlinear regime. All of this information is vital for optimization of the acceleration performance in plasma channel.

Finally, the saturation of the Raman backscattering spectrum provides information about the self-focusing conditions. As shown in Fig.\ref{fig2}a, the Raman based density curve starts to deviate from the straight line at the actual interferometry-based peak plasma density of $\sim 7 \times 10^{19} cm^{-3}$. Increasing the plasma density further, shifts the onset of self-focusing to lower densities, to a location in front of the density peak. By comparing the saturated Raman based plasma density with the real one, it is possible to estimate the density and/or laser intensity at which self-focusing initiates. Further, in designing  plasma structures to operate in linear to quasi-linear  LWFA regimes, saturation of the RBS spectrum can indicate the highest acceptable plasma density, beyond which the acceleration enters the bubble regime.

\section{Conclusion}

Raman backscattering was investigated experimentally and compared to analytical results in the weakly nonlinear approximation. Experiments were performed for different plasma densities, plasma shapes, laser pulse durations and intensities. The experimental results and nonlinear theory were in excellent agreement without the need for fitting parameters. The results demonstrate that at high pump intensities, the linear matching condition of Raman backscattering cannot be used to determine the plasma density from the Stokes wavelength. The numerical solution of the dispersion relation or the practical evaluation formula derived therefrom must be used instead. In general, the Raman backscattered wavelength is found to be a function of both plasma density and laser intensity. Knowledge of that function provides a novel diagnostic tool for laser-plasma interactions.

Most of the experiments reported here were performed in a stable laser wakefield accelerator operated in the density down-ramp bubble regime. Saturation of the nonlinear Raman backscattering spectrum was found to be correlated with electron acceleration, and explained by relativistic self-focusing. The self-focusing led to the destruction of the periodic structure of the plasma wave because of bubble and sheath formation, and wave breaking and phase-mixing processes that take place during LWFA. Saturation of the nonlinear Raman backscattering spectrum can be used as self-focusing diagnostic.

This work was supported by the NRL base program.


\begin{thebibliography}{10}

\bibitem{kruer1988}
W.L. Kruer, The Physics of Laser Plasma Interactions (Addison-Wesley, Redwood City, CA, 1988).

\bibitem{mc1992}
C.J. McKinstrie and R. Bingham, Stimulated Raman forward scattering and the relativistic modulational instability of light waves in rarefied plasma, Phys. Fluids B {\bf4}, 2626 (1992).

\bibitem{antonsen1993}
T.M. Antonsen and P. Mora, Self‐focusing and Raman scattering of laser pulses in tenuous plasmas, Phys. Fluids B 5, 1440 (1993).

\bibitem{gordon2002} 
D.F. Gordon, B. Hafizi, R.F. Hubbard and P. Sprangle, Raman sidescatter in numerical models of short pulse laser plasma interactions, Phys. Plasmas 9, 1157 (2002).

\bibitem{esarey2009}
E. Esarey, C.B. Schroeder and W.P. Leemans, Physics of laser-driven plasma-based electron accelerators, Rev. Mod. Phys. 81, 1229 (2009).

\bibitem{palastro2011}
J.P. Palastro, T.M. Antonsen Jr., A. Pearson, W. Zhu, and N. Jain, Raman scattering of intense, short laser pulses in modulated plasmas, Phys. Rev. E 83, 046410 (2011).

\bibitem{palastro2009}
J.P. Palastro, E.A. Williams, D.E. Hinkel, L. Divol, and D. Strozzi, Kinetic dispersion of Langmuir waves. I. The Langmuir decay instability, Phys. Plasmas 16, 092304 (2009).

\bibitem{Yin2008}
L. Yin, B.J. Albright, K.J. Bowers, W. Daughton, and H.A. Rose, Saturation of backward stimulated scattering of laser in kinetic regime: Wavefront bowing, trapped particle modulational instability, and traped particle self-focusing of plasma waves, Phys. Plasmas 15, 013109 (2008).

\bibitem{perry1992}
M. D. Perry, C. Darrow, C. Coverdale, and J. K. Crane, Measurement of the local electron density by means of stimulated Raman scattering in a laser-produced gas jet plasma, Optics Letters, 17, 523, (1992)

\bibitem{krushelnick1995}
K. Krushelnick, A. Ting, H. R. Burris,  A. Fisher, C. Manka, and E. Esarey, Second Harmonic Generation of Stimulated Raman Scattered Light in Underdense Plasmas, PRL 75, 3681 (1995).

\bibitem{jones2002}
T.G. Jones, K. Krushelnick, A. Ting, D. Kaganovich, C.I. Moore, A. Morozov, Temporally resolved Raman backscattering diagnostic of high intensity laser channeling, Review of Scientific Instruments, 73 (6), p. 2259 (2002).

\bibitem{ping2003}
Y. Ping, I. Geltner, and S. Suckewer, Raman backscattering (RBS) and amplification in a gas jet plasma, Phys. Rev. E, 67, 016401 (2003)

\bibitem{gordon2013}
D. F. Gordon, M. H. Helle, D. Kaganovich, A. Ting, and B. Hafizi, Laser Acceleration and Injection of Particles in Optically Shaped Gas Targets, Proc. of SPIE Vol. 8779, 877902 (2013).

\bibitem{kaganovich2014} 
D. Kaganovich, D. F. Gordon, M. H. Helle, and A. Ting, Shaping gas jet plasma density profile by laser generated shock waves, Journal of Applied Physics 116, 013304 (2014).

\bibitem{brantov2008}
A. V. Brantov, T. Zh. Esirkepov, M. Kando, H. Kotaki, V. Yu. Bychenkov, and S. V. Bulanov, Controlled electron injection into the wake wave using plasma density inhomogeneity, Phys. Plasmas 15, 073111 (2008).

\bibitem{helle2010}
M. H. Helle, D. Kaganovich, D. F. Gordon, and A. Ting, Measurement of Electro-Optic Shock and Electron Acceleration in a Strongly Cavitated Laser Wakefield Accelerator, Phys. Rev. Lett. 105, 105001 (2010).
 
\bibitem{kaganovich2013}
D. Kaganovich, J. R. Penano, M. H. Helle, D. F. Gordon, B. Hafizi, and A. Ting, Origin and control of the subpicosecond pedestal in femtosecond laser systems, Optics Letters 38, 3635 (2013).

\bibitem{esarey1997}
E. Esarey, P. Sprangle, J. Krall, and A. Ting, Self-focusing and guiding of short laser pulses in ionizing gases and plasmas, IEEE Trans. Plasma Sci. 33, 1879 (1997).

\bibitem{pukhov2002}
A. Pukhov and J. Meyer-ter-Vehn, Laser wake field acceleration: the highly non-linear broken-wave regime, Appl. Phys. B, 74, 355 (2002).
 
\bibitem{hafizi2000}
B. Hafizi, A. Ting, P. Sprangle and R. F. Hubbard, Relativistic focusing and ponderomotive channeling of intense laser beams, Phys. Rev. E 62, 4120 (2000).

\bibitem{kaganovich2015}
D. Kaganovich, J. P. Palastro, Y.-H. Chen, D. F. Gordon, M. H. Helle, A. Ting, Simulation of free-space optical guiding structure based on colliding gas flows, Applied Optics, 54, F144, (2015).

\bibitem{ehrlich1998}
Y. Ehrlich, C. Cohen, D. Kaganovich, A. Zigler, R. F. Hubbard, P. Sprangle, and E. Esarey, Guiding and damping of high-intensity laser pulses in long plasma channels, J. Opt. Soc. Am. 15, 2416 (1998).

\bibitem{hubbard2001}
R. F. Hubbard, D. Kaganovich, B. Hafizi, C. I. Moore, P. Sprangle, A. Ting, and A. Zigler, Simulation and design of stable channel-guided laser wakefield accelerators, Physical Review E, 63, 036502 (2001).

\end{thebibliography}
\end{document}